\documentclass[twocolumn,aps,prd,10pt]{revtex4-1}

\usepackage{amsmath}
\usepackage{amssymb}
\usepackage{amsfonts}
\usepackage{amsthm}
\usepackage{mathrsfs}

\newcommand{\bT}{\bar{T}}
\newcommand{\ditto}{''}
\newcommand{\SC}{\mathrm{SC}}
\newcommand{\GW}{\mathrm{GW}}
\newcommand{\EM}{\mathrm{EM}}

\newcommand{\bra}[1]{\langle#1|}

\newcommand{\ket}[1]{|#1\rangle}

\newcommand{\xx}{\boldsymbol{x}}
\newcommand{\kk}{\boldsymbol{k}}
\newcommand{\KK}{\boldsymbol{K}}

\newcommand{\ppp}{\\[3pt]}
\newcommand{\pppp}{\\[6pt]}
\newcommand{\nppp}{\nonumber\\[3pt]}
\newcommand{\npppp}{\nonumber\\[6pt]}
\newcommand{\LL}{{\mathcal L}}

\newcommand{\dd}{\mathrm{d}}
\newcommand{\TT}{\mathrm{TT}}
\newcommand{\Int}[1]{\substack{{\displaystyle\int}\\{#1}}}

\begin{document}

\title{Spacetime foam induced collective bundling of intense fields}

\author{Teodora Oniga}
\email{t.oniga@abdn.ac.uk}

\author{Charles H.-T. Wang}
\email{c.wang@abdn.ac.uk}
\affiliation{Department of Physics,
University of Aberdeen, King's College, Aberdeen AB24 3UE, United Kingdom}

%\date{}

\begin{abstract}
The influence of spacetime foam on a broad class of bosonic fields with arbitrary numbers of particles in the low-energy regime is investigated. Based on a recently formulated general description of open quantum gravitational systems, we analyse the propagation of scalar, electromagnetic, and gravitational waves on both long and short time scales with respect to their mean frequencies. For the long time propagation, the Markov approximation that neglects the effects of initial conditions of these waves is employed. In this case, despite intuitively expected decoherence and dissipation from the noisy spacetime, we show that such phenomena turn out to be completely suppressed for scalar bosons, photons, and gravitons, which are coupled to gravity but otherwise free. The short time effects are then recovered through the transient non-Markovian evolution. Focusing on scalar bosons in initially incoherent states, we find that the resulting quantum dissipation depends strongly on the distribution of the particle momentum states. We further identify a hitherto undiscovered collective anti-dissipation mechanism for a large number of  particles. The surprising new effect tends to ``bundle'' identical particles within sharply distributed momentum states having a width inversely proportional to the particle number due to the thermal fluctuations, or its square root due to the vacuum fluctuations of spacetime.
\end{abstract}

%\pacs{04.60.-m, 04.60.Bc, 03.65.Yz, 04.20.Cv}

% Research Areas: Quantum gravity, Open quantum systems & decoherence

\maketitle

\section{Introduction}

Decoherence is fundamental to quantum gravity phenomenology, as no physical systems are truly isolated from the gravitational environment. The existence of spacetime fluctuations thus provides a universal reservoir of environmental degrees of freedom with which all matter interacts, leading to the loss of coherence between superposition states. It has previously been suggested that through this mechanism, quantum gravity can have significant effects on quantum systems, in particular by causing the appearance of classical behaviour of matter and spacetime \cite{kiefer2, Giulini, Schlosshauer}. Furthermore, it imposes an ultimate limit on the type of precision measurements that can be performed due to the suppression of quantum interferences \cite{lamine, amelino, schiller}.

Environmental gravitational decoherence has been considered in the literature through a number of models  \cite{wang, ford1, breuer1, kay, anastopouloshu, blencowe, power, anastopoulos, Oniga2016}. These works have shown that gravitationally induced decoherence would cause generally small effects on laboratory scales \cite{Oniga2016, blencowe, Chen2014}. However, the issue of quantum systems of a large number of particles that could have different conclusions does not appear to be sufficiently addressed. While the models presented in Refs. \cite{blencowe,anastopoulos} would in principle allow for the treatment of many particles, only one-particle master equations have been studied in detail. Investigating collective decoherence and dissipation due to gravity could lead to unexpected effects, as even for noninteracting quantum particles, the existence of a common reservoir can enhance or suppress decoherence and dissipation \cite{palma}. In addition, the indistinguishable nature of the particles may contribute to such phenomena. Recently, we have derived a master equation for the decoherence and dissipation of general bosonic matter due to spacetime fluctuations \cite{Oniga2016}. Thanks to its wide versatility, we apply the generic formalism in this work to systems with an arbitrary number of identical particles and report on the resulting new collective quantum behaviours that may bear important physical consequences.

We use the relativistic units in which the speed of light, gravitational constant, and reduced Planck constant are, respectively, scaled as $c\to c/c=1$, $G\to G c^4$, and $\hbar\to \hbar/c$. This leaves two independent dimensions for physical quantities related to the
Planck length $\ell_\mathrm{P}$ and energy $E_\mathrm{P}$. Spatial components are enumerated with Latin indices with $1,2,3$.
Spacetime coordinates are labelled with Greek indices carrying an additional zero for time.
Shorthand $kx = k_\mu x^\mu$ is understood and summation over repeated coordinate and polarization indices is implied.
The time derivative and Hermitian conjugate are denoted with an over-dot $(\,\dot{}\,)$ and superscript dagger $(^\dag)$ respectively.
A ditto mark $(\ditto)$ refers to the previous expression or the right-hand side of the previous equation.

\section{Propagation of bosonic fields in fluctuating spacetime}
\label{sec2}

We start by considering a general bosonic field in an equilibrium bath of spacetime fluctuations characterised by a Gaussian state density matrix $\rho_\mathrm{B}$ of environmental gravitons. Unless otherwise stated, we assume the Planck distribution
%\begin{eqnarray}
$
%N_T(\omega)=\frac1{e^{\hbar\omega/k_\mathrm{B} T}-1}
N_T(\omega)=1/(e^{\hbar\omega/k_\mathrm{B} T}-1)
$
%\label{NT}
%\end{eqnarray}
at temperature $T$ is maintained, where $\omega$ is the graviton frequency and $k_\mathrm{B}$ is Boltzmann's constant.
%1

The statistical state of a generally non-equilibrium bosonic field is described by its density matrix $\rho(t)$ at time $t$ reduced from the total system as the ensemble average over the gravitational reservoir. Initially at $t=0$, this state is untangled with the environment so that the total state takes the factored form $\rho(0) \otimes \rho_\mathrm{B}$. Thereafter entanglements with the environment are developed leading to the reduced state, described in the interaction picture, by the following master equation
\begin{eqnarray}
&&\!\!\!\!\!\!\!\!\!\!\!\!\!\!
\dot\rho(t)
=
-\frac{i}{\hbar} [H_\mathrm{SG}, \rho(t)]
-\frac{8\pi G}{\hbar}
\int\!\! \frac{\dd^3 k}{2(2\pi)^3k}
\nppp&&
\times\,
\Big \{
\int_{0}^{t} \!\!\dd t'
e^{-i k (t - t')}
\big(
[
\tau^\dag_{ij} (\kk, t),\,
\tau_{ij}(\kk, t') \rho(t)
]
\nppp
&&
\,+\,N_T(k)\,
[
\tau^\dag_{ij} (\kk, t),\,
[
\tau_{ij}(\kk, t'),\,
\rho(t)
]]
\big)
+(\ditto)^\dag
\Big\}
\label{maseqn}
\end{eqnarray}
recently derived in Ref. \cite{Oniga2016},
where
\begin{eqnarray*}
H_\mathrm{SG}
=
-2G\!\int\!\dd^3 x\,\dd^3 x'
\frac{T^{\mu\nu}(\xx,t)\bT_{\mu\nu}(\xx',t-|\xx-\xx'|)}
{|\xx-\xx'|}
%\label{HIU}
\end{eqnarray*}
denotes the self-gravity Hamiltonian and
\begin{eqnarray}
\tau_{ij}(\kk, t)
&=&
\int\!\dd^3x \,
\tau_{ij}(\xx, t)
e^{-i \kk\cdot\xx}
\end{eqnarray}
in terms of the stress-energy tensor $T_{\mu\nu}$, its trace-reversion $\bT_{\mu\nu}$ and transverse-traceless (TT) stress tensor $\tau_{ij}=T^\TT_{ij}$ of the field.

For a scalar (SC) field with dispersion relation
$\omega_k^2 = k^2 + \mu^2$
using a mass parameter $\mu \ge 0$ and annihilation $a_{\kk}$ and creation $a_{\kk}^\dag$ operators satisfying
\begin{eqnarray}
[a_{\kk}, a_{\kk'}^\dag]
=
\delta^3(\kk-\kk'),
%\label{com1}
%\nppp
\;
[a_{\kk}, a_{\kk'}]
&=&
0\,=\,
[a_{\kk}^\dag, a_{\kk'}^\dag]
\label{com3}
\end{eqnarray}
we find the corresponding TT stress tensor operator to be
\begin{eqnarray}
&&\hspace{-20pt}
\tau_{ij}^\SC(\kk,t)
=
%-\frac{\hbar}{2}\, P_{ijkl}(\kk)\!\int\dd^3k' % wrong sign!!
\frac{\hbar}{2}\, P_{ijkl}(\kk)\!\int\dd^3k'
\frac{k'_k k'_l}{\sqrt{\omega_{\kk'}}}
\nppp&&
\times\,
\frac{a^\dag_{\kk'-\kk} a_{\kk'} e^{-i(\omega_{\kk'}-\omega_{\kk'-\kk})t}}
{\sqrt{\omega_{\kk'-\kk}}}
+
(\ditto|_{\kk\to-\kk})^\dag
\label{stij1}
\end{eqnarray}
where
$P_{ijkl}=(P_{ik}P_{jl} + P_{il}P_{jk} -  P_{ij}P_{kl})/2$
is the TT projector in terms of the transverse projector $P_{ij}$ \cite{Flanagan2005}.
In arriving at the above $\tau_{ij}^\SC(\kk,t)$, we have adopted normal ordering and retained terms with particle number preserving matrix elements to be consistent with the low-energy effective open quantum system description that involves no pair productions or destructions.

For the electromagnetic (EM) field, in a fashion analogous to Eq. \eqref{stij1}, we find the Maxwell TT stress tensor to be
\begin{eqnarray}
&&\hspace{-15pt}
\tau_{ij}^\EM(\kk,t)
=
-\frac{\hbar}{2}\, P_{ijkl}(\kk)\!\int\dd^3k'
\sqrt{\omega_{\kk'-\kk}\omega_{\kk'}}
\nppp&&
\times\,
\big[
e_{k}^{\sigma}(\kk'-\kk )\, e_{l}^{\upsilon}(\kk')
+
\epsilon_{\sigma\sigma'}
\epsilon_{\upsilon\upsilon'}
e_{k}^{\sigma'}(\kk'-\kk)\,
e_{l}^{\upsilon'}(\kk')
\big]
\nppp&&
\times\,
\frac{a^{\sigma\dag}_{\kk'-\kk} a^{\upsilon}_{\kk'} e^{-i(\omega_{\kk'}-\omega_{\kk'-\kk})t}}
{\sqrt{\omega_{\kk'-\kk}}}
+
(\ditto|_{\kk\to-\kk})^\dag
\label{etij1}
\end{eqnarray}
in terms of the basis polarisation vectors $e_{i}^{\sigma}(\kk)$ for $\sigma=1,2$,
normalised with $e_{i}^\sigma(\kk)e_{j}^\sigma(\kk)=P_{ij}(\kk)$,
and the annihilation $a_{\kk}^{\sigma}$ and creation $a_{\kk}^{\sigma}{}^\dag$ operators for photons satisfying similar commutation relations \eqref{com3} with additional orthogonality between two independent polarisations.

Similar treatment can be extended to the non-equilibrium gravitational field considered as a bosonic system subject to environmental decoherence and dissipation caused by equilibrium gravitational fluctuations. For this purpose, there is a need to distinguish these two contributions to the metric perturbations. We thus denote the two types of metric perturbations  by $h_{\mu\nu}$ due to the stronger non-environmental gravitational wave (GW) and by $q_{\mu\nu}$ due to the weaker environmental metric fluctuations which also accounts for the self-gravity of the GW. This then  leads to spacetime metric of the form
%\begin{eqnarray}
$g_{\mu\nu} = \eta_{\mu\nu} + h_{\mu\nu} + q_{\mu\nu}$,
%\label{hq}
%\end{eqnarray}
where $\eta_{\mu\nu} = \mathrm{diag}(-1,1,1,1)$ is the background Minkowski metric, and the perturbation orders for $h_{\mu\nu}$ and $q_{\mu\nu}$ are considered to start from the first and second orders, respectively.
It is important to note that for the perturbative metric expansion to be physically consistent, the energy density of the gravitational source must be below the Planck scale \cite{Anderson2003}.

Up to the second-order perturbations, they satisfy the effective Einstein equations
\begin{eqnarray}
G_{\mu\nu}^{(1)}[h_{\alpha\beta}]
= 0,\;
%\label{G1}
%\ppp
G_{\mu\nu}^{(1)}[q_{\alpha\beta}]
=
-G^{(2)}_{\mu\nu}[h_{\alpha\beta}]
\label{G12}
\end{eqnarray}
where
\begin{eqnarray*}
G_{\mu\nu}^{(1)}[h_{\alpha\beta}]
&=&
\frac12\big\{
h^{\mu\alpha}{}_{,}{}^\beta{}_\mu
+h^{\mu\beta}{}_{,}{}^\alpha{}_\mu
-h^{\alpha\beta}{}_{,}{}^\mu{}_\mu
\npppp&&
-h^\nu{}_{\nu,}{}^{\alpha\beta}
+\eta^{\alpha\beta}h^\nu{}_{\nu,}{}^\mu{}_\mu
-\eta^{\alpha\beta}h^{\mu\nu}{}_{,\mu\nu}
\big\}
\ppp
G_{\mu\nu}^{(2)}[h_{\alpha\beta}]
&=&
\big\{
R_{\mu\nu}^{(2)}
- \frac12 \eta_{\mu\nu}R^{(2)}
- \frac12 h_{\mu\nu} R_{\mu\nu}^{(1)}
\big\}[h_{\alpha\beta}]
\end{eqnarray*}
are the first and second order expansions of the Einstein tensor,
in terms of the first and second order expansions  $R_{\mu\nu}^{(1)}$ and $R_{\mu\nu}^{(2)}$ of the Ricci tensor  respectively
\cite{MTW1973, Flanagan2005}.

It is interesting to note that equations \eqref{G12} above naturally arise from the Lagrangian density
\begin{eqnarray}
\LL &=& \frac{1}{16\pi G}
\big\{
R^{(2)}[h_{\alpha\beta}] + R^{(2)}[q_{\alpha\beta}]
- q_{\mu\nu} G^{(2)\mu\nu}[h_{\alpha\beta}]
\big\}
\nppp
\label{LL2}
\end{eqnarray}
as variational field equations for $h_{\mu\nu}$ and $q_{\mu\nu}$.
This Lagrangian density $\LL$ fits the form constituting an open quantum gravitational system \cite{Oniga2016}, with the GW field $h_{\mu\nu}$ acting as an effective bosonic field, whereas the inhomogeneous and homogeneous solutions for $q_{\mu\nu}$ describe the self-gravity of $h_{\mu\nu}$ and spacetime fluctuations respectively.

Henceforth, by gravitons we will primarily relate to $h_{\mu\nu}$ having an effective stress-energy tensor, from the preceding discussion, given by
%\begin{eqnarray}
$
T_{\mu\nu}^\GW
=
-\frac1{8\pi G}\,G^{(2)}_{\mu\nu}[h_{\alpha\beta}]
$.
To proceed, we apply the TT condition on $h_{ij}=h_{ij}^\TT$ and obtain from the TT part of $T_{\mu\nu}^\GW$
\begin{eqnarray}
\tau_{ij}^\GW
&=&
\frac1{16\pi G}
P_{ijkl}
\big[
2h_{mn}h_{km,ln}
+h_{km,n}h_{ln,m}
\nppp
&&%\hspace{-10pt}
-h_{km,n}h_{lm,n}\!
-\frac12h_{mn,k}h_{mn,l}
\big].
\label{qqtau2}
\end{eqnarray}
The quantised GW field $h_{ij}$ has the expansion
\begin{eqnarray}
h_{ij}
=
\int\!\dd^3 k
%\sqrt{\frac{16\pi G \hbar}{2(2\pi)^3 k}}\;
\sqrt{\frac{8\pi G \hbar}{(2\pi)^3 k}}\;
e_{ij}^\sigma(\kk)\,
a_{\kk}^\sigma\,
e^{ikx}
+
(\ditto)^\dag
%\label{gwex}
\end{eqnarray}
in terms of the basis TT tensors $e_{ij}^\sigma(\kk)$ for $\sigma=1,2$,
normalised with $e_{ij}^\sigma(\kk)e_{kl}^\sigma(\kk)=2\,P_{ijkl}(\kk)$,
and the annihilation $a_{\kk}^{\sigma}$ and creation $a_{\kk}^{\sigma}{}^\dag$ operators for gravitons satisfying similar commutation relations for photons.
Substituting this expansion of $h_{ij}$  into \eqref{qqtau2}, we can readily obtain the normal ordered and particle number preserving operator $\tau_{ij}^\GW(\kk,t)$, similar in form to \eqref{etij1} for photons, involving wave vector $\kk$ as well as polarisation tensors $e_{ij}^\sigma(\kk)$ instead of  polarisation vectors $e_{i}^\sigma(\kk)$.

To investigate the non-unitary evolution of these bosonic fields in the momentum basis, it is convenient to express \eqref{stij1} in the form
\begin{eqnarray}
\!\!
\tau_{ij}^\SC(\kk,t)
=
\int\!\dd^3k' \tau_{ij}^\SC(\kk,\kk') e^{-i\omega_{\kk,\kk'} t}
+
(\ditto|_{\kk\to-\kk})^\dag
\label{tkksc}
\end{eqnarray}
in terms of
$
\omega_{\kk,\kk'}=\omega_{\kk'}-\omega_{\kk'-\kk}
$
and
\begin{eqnarray}
&&\hspace{-20pt}
\tau_{ij}^\SC(\kk,\kk')
=
%-\frac{\hbar}{2}\, P_{ijkl}(\kk) % wrong sign
\frac{\hbar}{2}\, P_{ijkl}(\kk)
\frac{k'_k k'_l}{\sqrt{\omega_{\kk'}}}
\frac{a^\dag_{\kk'-\kk} a_{\kk'}}
{\sqrt{\omega_{\kk'-\kk}}} .
\label{stij1kk}
\end{eqnarray}
Analogously, we can write
\begin{eqnarray}
\tau_{ij}(\kk,t)
=
\int\!\dd^3k' \tau_{ij}(\kk,\kk') e^{-i\omega_{\kk,\kk'} t}
+
(\ditto|_{\kk\to-\kk})^\dag
\label{tkk}
\end{eqnarray}
with corresponding constructions for $\tau_{ij}=\tau_{ij}^\SC$, $\tau_{ij}^\EM$ and $\tau_{ij}^\GW$.
Furthermore, using the orthogonality relations between these quantities, we can derive a useful property,
\begin{eqnarray}
\tau_{ij}(\kk,\mathrm{const}\times\kk)=0
\label{tkkvsh}
\end{eqnarray}
valid also for scalar bosons, photons, and gravitons with $\tau_{ij} = \tau_{ij}^\SC$, $\tau_{ij}^\EM$, and $\tau_{ij}^\GW$ respectively.

Using the properties developed so far in this section, we can now establish the long-term influence of spacetime fluctuations on the propagation of the above bosonic fields in terms of decoherence and dissipation. For this purpose, let us carry out the Markov approximation \cite{Breuer2002} of the general master equation \eqref{maseqn} by changing the time variable $t'=t-s$, taking the limit
%\begin{eqnarray}
$
\int_{0}^{t} \dd s
\to
\int_{0}^{\infty} \dd s
$
%\end{eqnarray}
corresponding to infinitely long past history,
and applying the Sokhotski-Plemelj theorem
%\begin{equation}
$
\int_0^\infty \dd s\, e^{-i \epsilon s}
= \pi \delta(\epsilon) - i \,\mathrm{P} \frac{1}{\epsilon}
$
%\end{equation}
where $\mathrm{P}$ denotes the Cauchy principal value that gives rise to a non-dissipative Hamiltonian $H_\mathrm{LS}$ for possible Lamb and Stark shifts \cite{Breuer2002}. Together with Eq. \eqref{tkk}, this process allows the master equation \eqref{maseqn} to reduce to the Markovian form
\begin{eqnarray}
&&\!\!\!\!\!
\dot\rho(t)
=
-\frac{i}{\hbar} [H_\mathrm{SG}+H_\mathrm{LS}, \rho(t)]
-\frac{8\pi^2 G}{\hbar}
\int \frac{\dd^3 k\,\dd^3 k'}{2(2\pi)^3k}\sum_{\pm}
\nppp&&\!
\times\,
\delta(k \mp \omega_{\pm\kk,\kk'})
\Big\{
e^{\mp i\omega_{\pm\kk,\kk'} t}
\big(
[
\tau^\dag_{ij} (\kk, t),\,
\tau_{ij}^\pm(\kk, \kk') \rho(t)
]
\nppp
&&\!
+\,N_T(k)\,
[
\tau^\dag_{ij} (\kk, t),\,
[
\tau_{ij}^\pm(\kk, \kk'),\,
\rho(t)
]]
\big)
+(\ditto)^\dag
\Big\}
\label{maseqn2}
\end{eqnarray}
with $\tau_{ij}^+(\kk, \kk')=\tau_{ij}(\kk, \kk')$,
$\tau_{ij}^-(\kk, \kk')=\tau_{ij}^\dag(-\kk, \kk')$, and
$\sum_{\pm}$ indicating the two expressions with corresponding alternative signs are summed together. We will refer to the non-unitary part of \eqref{maseqn2} as the Markov dissipator.

From Eq. \eqref{stij1} we see that the nonzero condition $k=\pm\omega_{\pm\kk,\kk'}$ for the $\delta$-function in Eq. \eqref{maseqn2} requires the field to be massless and wave vectors $\kk$ and $\kk'$ to be parallel. But this makes $\tau_{ij}(\kk,\kk') =0$ according to Eq. \eqref{tkkvsh} for ``free'' scalar bosons, photons and gravitons in the sense used in this work that they interact only with gravitation and are otherwise free. Consequently, for these particles, the Markov dissipator in \eqref{maseqn2} vanishes, resulting in the lack of long-term quantum decoherence and dissipation due to spacetime fluctuations \cite{endnote}.

\section{Collective bundling of identical bosons by spacetime foam}

To obtain nontrivial quantum dissipative effects during the transient evolution of an ensemble $\rho(t)$ of $N$ particles from an initial state $\rho(0)$ untangled with the gravitational reservoir, we return to the master equation \eqref{maseqn}
and consider these unbound particles to have a sufficient spatial spread for their self-gravity to be negligible by approximating
%\begin{eqnarray}
$
[H_\mathrm{SG}, \rho(t)]=0
$.
%\label{SG0}
%\end{eqnarray}

Given the complex dynamical structures arising from the dominant non-Markovian dissipator in Eq. \eqref{maseqn}, let us champion the development with relatively simple scalar bosons that  have been found to share a number of key properties with photons and gravitons in the previous section.
The underlying $N$-particle Hilbert space is spanned by states of the following generic form
\begin{eqnarray}
\ket{\KK}
=
\frac{1}{\sqrt{N!}}\,a_{\kk^1}^\dag \cdots a_{\kk^N}^\dag\ket{0}
%\label{psi}
\end{eqnarray}
where $\KK = \{\kk^1,\kk^2,\cdots,\kk^N\}$ denotes a set of  wave vectors of the $N$ particles.

Even with scalar bosons as our ``explorer'' particles using the TT stress tensor $\tau_{ij} = \tau_{ij}^\SC$, the evaluation of the matrix elements $\bra{\KK}\cdot\ket{\KK'}$ of the non-Markovian dissipator with respect to general $\ket{\KK}$ and $\ket{\KK'}$ can  still be quite involved. Fortunately, by exploiting a number of identities, we are able to gain considerable simplifications resulting in a trackable approach.
To begin with, we notice from the normal ordering and particle number conservation of $\tau_{ij}$ that
%\begin{eqnarray}
$
\tau_{ij}(\kk,t)\ket{0}=0
$
%=\tau_{ij}^\dag(\kk,t)\ket{0}
%\label{tau0}
%\end{eqnarray}
and from Eq. \eqref{stij1} and indeed more generally Eq. \eqref{tkk}, we also have
%\begin{eqnarray}
$
\tau_{ij}(-\kk,t)
=
\tau_{ij}^\dag(\kk,t)
$.
%\end{eqnarray}
These properties enable us to infer the (multiple) actions of the TT stress tensor operators on an $N$-particle state from the commutation relation
\begin{eqnarray}
[\tau_{ij}(\kk,t),a_{\kk'}]
=
%\hbar P_{ijkl}(\kk) % wrong sign!!
-\hbar P_{ijkl}(\kk)
\frac{k'_k k'_l e^{-i(\omega_{\kk'+\kk}-\omega_{\kk'})t}}
{\sqrt{\omega_{\kk'+\kk}\omega_{\kk'}}}
a_{\kk'+\kk}
\nppp
\label{ta1}
\end{eqnarray}
%\begin{eqnarray}
%&&\hspace{-30pt}
%[\tau_{ij}(\kk,t),a_{\kk'}]
%\nppp
%&=&
%\hbar P_{ijkl}(\kk) k'_k k'_l
%\frac{e^{-i(\omega_{\kk'+\kk}-\omega_{\kk'})t}}
%{\sqrt{\omega_{\kk'+\kk}\omega_{\kk'}}}
%a_{\kk'+\kk}
%\label{ta1}
%\end{eqnarray}
and other commutators between
$\{\tau_{ij}(\kk,t),\tau_{ij}^\dag(\kk,t)\}$
and
$\{a_{\kk'}, a_{\kk'}^\dag\}$
obtained upon taking $(\kk\to\pm\kk)^{(\dag\,)}$ of \eqref{ta1}.
These commutators can be regarded as super-operators on $a_{\kk'}, a_{\kk'}^\dag$ that shift their wave vectors by $\pm\kk$.

To evaluate the double commutators in the master equation~\eqref{maseqn}, we proceed in the same manner by successively applying similar super-operators with compensating shifts in momenta. This leads to further eigen super-operator equations of the following form
\begin{eqnarray}
&&\hspace{-30pt}
[\tau_{ij}(\kk,t'),[\tau_{ij}^\dag(\kk,t), a_{\kk'}]]
\nppp
&=&
\hbar^2 P_{ijkl}(\kk) k'_i k'_j k'_k k'_l
\frac{e^{-i(\omega_{\kk'-\kk}-\omega_{\kk'})(t-t')}}
{{\omega_{\kk'-\kk}\omega_{\kk'}}}
\,a_{\kk'} .
\label{tad1}
\end{eqnarray}
Relations exemplified by Eqs. \eqref{ta1}, \eqref{tad1} and their extensions with $a_{\kk'}^{(\dag\,)} \to a_{\kk^1}^{(\dag\,)} \cdots a_{\kk^N}^{(\dag\,)}$ provide beneficial tools to systematically evaluate all the matrix elements
$\bra{\KK}\cdot\ket{\KK'}$
of the non-Markovian dissipator of the master equation~\eqref{maseqn}. The resulting matrix elements with $\ket{\KK}=\ket{\KK'}$ and $\ket{\KK}\neq\ket{\KK'}$ are responsible for dissipation and decoherence respectively, each having its own rich dynamical content.

Below we will focus on dissipation and demonstrate its significant effects, leaving more detailed study on the $N$-particle decoherence using \eqref{maseqn} to future work. Transient dissipation of an incoherent statistical state of the $N$-particle scalar system can play a significant role for realistic sources that are classical in nature. Here, its analysis is based on a further available observation: if the density matrix $\rho(t)$ is initially diagonal in the momentum basis, then under quantum evolution using the master equation \eqref{maseqn}, density matrix $\rho(t)$ remains diagonal in the same basis.

To see this, consider a diagonal density matrix $\rho(t)$ with elements
$\bra{\KK}\rho(t)\ket{\KK} = f(\KK,t)$
using a real function $f$ depending symmetrically on the particle wave vectors.
This function is normalised with
%\begin{eqnarray}
$
\int f \;\dd^{3N}\!K =1
$
%\end{eqnarray}
to ensure the unity trace condition of the density matrix $\rho(t)$.
Using relations \eqref{ta1} and \eqref{tad1} and their extensions described above, we obtain after some algebra the following diagonal elements associated with the non-Markovian dissipator of Eq. \eqref{maseqn}:
\begin{eqnarray}
\bra{\KK}\tau^\dag_{ij} (\kk, t)\tau_{ij}(\kk, t')\rho(t)\ket{\KK}
&=&
\hbar^2 P_{ijkl}(\kk)
\sum_{I=1}^{N}
\nppp&&
\hspace{-110pt}\times\,k^I_{i} k^I_{j} k^I_{k} k^I_{l}
\frac{e^{-i(\omega_{\kk^I-\kk}-\omega_{\kk^I})(t-t')}}
{{\omega_{\kk^I-\kk}\omega_{\kk^I}}}
f(\KK,t)
\npppp
\bra{\KK} \rho(t) \tau_{ij}(\kk, t') \tau^\dag_{ij} (\kk, t) \ket{\KK}
&=&
(\ditto|_{\kk\to-\kk})^\dag ,
\label{npart1}
\end{eqnarray}
and
\begin{eqnarray}
\bra{\KK}\tau_{ij}(\kk, t')\rho(t)\tau^\dag_{ij} (\kk, t)\ket{\KK}
&=&
\hbar^2 P_{ijkl}(\kk)
\sum_{I=1}^{N}\sum_{J=1}^{N}
\nppp&&
\hspace{-110pt}\times\,k^I_{i} k^I_{j} k^J_{k} k^J_{l}
\frac{e^{i(\omega_{\kk^J+\kk}-\omega_{\kk^J})t}}
{\sqrt{\omega_{\kk^J+\kk}\omega_{\kk^J}}}
\frac{e^{-i(\omega_{\kk^I+\kk}-\omega_{\kk^I})t'}}
{\sqrt{\omega_{\kk^I+\kk}\omega_{\kk^I}}}
%\npppp&&
%\hspace{-110pt}\times\,
% f(\kk^1,\cdots,\kk^J+\kk,\cdots ,\kk^N;t)
\npppp&&
\hspace{-110pt}\times\,
 f(\KK |_{\kk^J\to\kk^J+\kk},t)
\npppp
\bra{\KK} \tau^\dag_{ij} (\kk, t) \rho(t) \tau_{ij}(\kk, t') \ket{\KK}
&=&
(\ditto|_{\kk\to-\kk})^\dag .
\label{npart2}
\end{eqnarray}
It is therefore clear that the master equation \eqref{maseqn} with negligible self-gravity transcends into a closed system under time evolution for function $f$, thereby preserving the stated diagonal structure of the incoherent density matrix $\rho$.

Some more striking observations can now be made, in the special case with $\ket{\KK}$ consisting of $N$ particles occupying the same momentum states, i.e. $\kk^1=\kk^2=\cdots=\kk^N$. Then while the single sum over particles in Eq. \eqref{npart1} contributes to positive dissipation proportional to $N$,
the double sum in Eq. \eqref{npart2} contributes to negative dissipation in proportion to $N^2$. On the other hand, unlike Eq. \eqref{npart1}, the $f$ factor in \eqref{npart2} depends on $\kk$, which in the presence of a dynamical width of the distribution of $f$ over $\kk$, would serve to limit such negative dissipation after integrating over $\kk$ in Eq. \eqref{maseqn}. These two competing factors suggest the existence of an equilibrium width of $\kk$ that decreases with the particle number $N$.

Specifically, let us assume $\rho(t)$ to have a characteristic width $\Delta k$ around a mean wave vector $\bar{\kk}$ initially. To be consistent with the negligible self-gravity for a large spatial spread of the $N$-particle system, the width $\Delta k$ should be limited, which will be taken to be $\Delta k < \bar{k}$ for simplicity. In accordance with the low-energy effective description of Eq. \eqref{maseqn}, when carrying out the integrations over $\kk$, we will adopt a UV cutoff naturally related to the energy scale of the $N$-particle system so that $k < \bar{k}$.

%Given $N$, this critical $\Delta k$ defines the width of bundling that a wider or narrower $\kk$-distribution of particle will tend towards.

With the above understanding, we evaluate the diagonal matrix element of master equation \eqref{maseqn}, using Eqs. \eqref{npart1} and \eqref{npart2}, related to state $\ket{\KK}$ consisting of $N$ particles all having the mean wave vector, i.e. $\kk^1=\kk^2=\cdots=\kk^N=\bar{\kk}$.
A transient time interval $\Delta t=2\pi/\bar{k}$ then emerges so that the corresponding matrix element of the non-Markovian dissipator approaches zero rapidly for $t > \Delta t$, which is consistent with the lack of Markovian dissipator in long-time evolution.

During the transient period $0 < t < \Delta t$, the density matrix element $\bra{\KK}\rho(t)\ket{\KK}$ abbreviated with $\varrho(t)$ in the following, changes in terms of vacuum
($\Delta\varrho_\mathrm{vac}$)
and thermal
($\Delta\varrho_T$)
contributions due to the first and second terms of the non-Markovian dissipator in Eq. \eqref{maseqn} respectively. Up to an order 1 factor, we find these changes to be given approximately by
\begin{eqnarray}
\!\!\!\!\!\!
\frac{\Delta\varrho_\mathrm{vac}}{\varrho(0)}
&\simeq&
\frac{3 N G \hbar \bar{k}^2}{2\,\omega_{\bar{\kk}}^2}
\Big(
N\!\Int{k<\Delta k}\! - \Int{k<\bar{k}}\,
\Big)
k\,\dd k ,
\label{disvac}
\pppp
\!\!\!\!\!\!
\frac{\Delta\varrho_T}{\varrho(0)}
&\simeq&
\frac{3 N G \hbar\bar{k}^2}{\omega_{\bar{\kk}}^2}
\Big(
N\!\Int{k<\Delta k}\! - \Int{k<\bar{k}}\,
\Big)
N_T(k)k\,\dd k .
\label{disth}
\end{eqnarray}
Evidently the above expressions can flip signs depending on the wave vector width $\Delta k$ and the particles number $N$, leading to a ``bundling'' effect with a characteristic width $\Delta k$ for each of Eqs. \eqref{disvac} and \eqref{disth} when their right-hand sides vanish. It is clear that in each case the width $\Delta k$ decreases with the particles number $N$.
Such a tendency for bosonic particles to gather towards the same state as a result of balanced positive and negative dissipations may be compared to the collective bunching effect for photons \cite{HBT1956} and other systems that appear to be classical in nature while satisfying Bose-Einstein statistics \cite{HBT1998}. Here we find this bosonic behaviour could be further enhanced by spacetime fluctuations.

Since fundamentally vacuum and thermal fluctuations of spacetime take place on scales of the Planck length $\ell_\mathrm{P}$ and thermal correlation length
%\begin{eqnarray}
$
\ell_T = \frac{\hbar}{k_\mathrm{B} T}
$
%\end{eqnarray}
at temperature $T$ respectively, it is instructive to analyse Eqs. \eqref{disvac} and \eqref{disth} in terms of these scales. The approximation $N_T(k)\simeq k_\mathrm{B} T/\hbar k$ will be used to explicitly evaluate \eqref{disth} in the low-energy limit with $\hbar\bar{k} < k_\mathrm{B} T$ as a definitive and physically reasonable scenario which may be readily generalised without affecting the general arguments here.

Then, for massless particles, Eqs. \eqref{disvac} and \eqref{disth} reduce, respectively, to
\begin{eqnarray}
\frac{\Delta\varrho_\mathrm{vac}}{\varrho(0)}
&\simeq&
\frac{3N \ell_\mathrm{P}^2}{4}
( N {\Delta k}^{\,2} - \bar{k}^{\,2} ) ,
\label{disvac3}
\ppp
\frac{\Delta\varrho_T}{\varrho(0)}
&\simeq&
\frac{3 N \ell_\mathrm{P}^2}{\ell_T}
( N{\Delta k} - \bar{k} ) .
\label{disth3}
\end{eqnarray}
In this case, using Eq. \eqref{disvac3}, the the bundling effect due to vacuum spacetime fluctuations has a momentum width given by
\begin{eqnarray}
\frac{\Delta p}{p} \simeq \frac{1}{\sqrt{N}}
\label{wdthvac}
\end{eqnarray}
where $p=\hbar\bar{k}$ is the mean particle momentum with $\Delta p=\hbar \Delta k$. For this effect to be effective during the transient period, or equivalently for significant bundling to occur, we require $\Delta\varrho_\mathrm{vac}/\varrho(0) > 1$ for an initial width $\Delta k \sim \bar{k}$. This implies the condition
\begin{eqnarray}
N > \frac{\lambda}{\ell_\mathrm{P}} \simeq \frac{E_\mathrm{P}}{E}
\label{Nvac}
\end{eqnarray}
where $\lambda=2\pi/\bar{k}$ is the mean wavelength of the particles,
on the number of particles, where $E$ is the mean energy of the particles. This condition amounts to the total energy carried by the $N$ particles exceeding the Planck energy.
However, for the perturbative metric expansion to be valid, the ratio of energy $E$ to cubic wavelength $\lambda^3$ must be so that the effective energy density is less than the Planck density:
%\begin{eqnarray}
$
\frac{E}{\lambda^3} < \frac{E_\mathrm{P}}{\ell_\mathrm{P}^3} .
$
%\label{pden}
%\end{eqnarray}
Otherwise non-perturbative quantum effects could also contribute and are beyond the scope of our present considerations.

Using Eq. \eqref{disth3}, the similar bundling effect due to thermal fluctuations of spacetime has a momentum width given by
\begin{eqnarray}
\frac{\Delta p}{p} \simeq \frac{1}{N}
\label{wdthth}
\end{eqnarray}
which can be substantially smaller with larger $N$ compared with the vacuum case in \eqref{wdthvac}.
Moreover, significant bundling could occur with $\Delta\varrho_{T}/\varrho(0) > 1$ if
\begin{eqnarray}
N > \frac{\sqrt{\ell_T\lambda}}{\ell_\mathrm{P}}.
%\simeq \frac{E_\mathrm{P}}{\sqrt{E_T E}}.
\label{Nth}
\end{eqnarray}
Compared with Eq.  \eqref{Nvac}, this condition is a less stringent condition on $N$ if the mean energy of the particles $E$ is less than the mean thermal energy of the environmental gravitons $\sim k_\mathrm{B} T$ with $\ell_T < \lambda$.

For nonrelativistic massive particles, Eqs. \eqref{disvac} and \eqref{disth} reduce, respectively, to
\begin{eqnarray}
\frac{\Delta\varrho_\mathrm{vac}}{\varrho(0)}
&\simeq&
\frac{3N \ell_\mathrm{P}^2}{4}
\frac{\bar{k}^2}{\mu^2}
( N {\Delta k}^{\,2} - \bar{k}^{\,2} ) ,
\label{disvacn}
\ppp
\frac{\Delta\varrho_T}{\varrho(0)}
&\simeq&
\frac{3 N \ell_\mathrm{P}^2}{\ell_T}
\frac{\bar{k}^2}{\mu^2}
( N{\Delta k} - \bar{k} ) .
\label{disthn}
\end{eqnarray}
Using Eq. \eqref{disvacn}, the bundling effect due to vacuum spacetime fluctuations also has a width given by Eq. \eqref{wdthvac} and this effect could be significant under the condition
\begin{eqnarray}
N
>
%\frac{\mu}{\ell_\mathrm{P}\bar{k}^2}
%\simeq
\frac{E_\mathrm{P}}{{E}_\mathrm{kin}}
\label{NNvac}
\end{eqnarray}
where ${E}_\mathrm{kin}$ is the mean Newtonian kinetic energy of the particles.
Again, using Eq. \eqref{disthn}, the bundling effect due to thermal spacetime fluctuations has a smaller width given by Eq. \eqref{wdthth}, with significant bundling if
\begin{eqnarray}
N
>
%\frac{\sqrt{\ell_T\bar\lambda}\,\mu}{\ell_\mathrm{P} \bar{k}}
%\simeq
\sqrt{\frac{\ell_T}{\lambda}}\,\frac{E_\mathrm{P}}{{E}_\mathrm{kin}}
\label{NNth}
\end{eqnarray}
is satisfied, where $\lambda$ is the mean de Broglie wavelength of the particles. One also sees that this condition is less stringent on $N$ compared with Eq. \eqref{NNvac}, if the mean de Broglie wavelength of the particles is longer than the correlation length of thermal environmental gravitons.

\newpage

\section{Concluding remarks}
\label{sec4}

For the first time we have described theoretically a novel collective quantum gravitational effect leading to an anti-dissipation mechanism that amplifies with the number $N$ of the identical particles. The effect is highly counterintuitive as spacetime foam phenomena, if any, have been widely anticipated to manifest some kind of randomisation of radiation and matter \cite{Amelino2013, Arteaga, Garay1998, Schafer1981}. Instead, with the exception of $N=1$ already reported in Ref. \cite{Oniga2016}, here we show that during the transient evolution from an initial state untangled with the gravitational environment, following emission or experimental preparation, identical particles could be bundled towards a spike, rather than scattered, in the momentum space by metric fluctuations.

Although this mechanicalism is presently demonstrated using a scalar field, other (pseudo) bosonic fields, such as electromagnetic and gravitational waves, are expected to behave likewise. Using Eq. \eqref{Nvac} for the vacuum contribution we see that such a scenario would be important for intense light sources or strong gravitational wave events similar to recent observation in Ref. \cite{LIGO2016}, where the total energy of the $N$-particle system around the mean frequency are likely to exceed the Planck energy. From Eq. \eqref{Nth} the thermal contribution also depends on the currently uncertain  temperature $T$ of the gravitational bath and could enhance the vacuum values based on $T \sim 1$ K and gravitational wavelength $\lambda \gg \ell_T \sim 1$ mm. In contrast, from Eqs. \eqref{NNvac}  and  \eqref{NNth} with also $T \sim 1$ K, this bundling effect could be less significant for ensembles of non-relativistic identical particles such as Bose-Einstein condensates under typical laboratory conditions.

Following the transient period, in which the new bundling mechanicalism is effective, we show that gravitationally interacting free particles experience no further gravitational decoherence and dissipation. This underlines a commonly unsuspected fact that spacetime foam by itself may not be enough to cause non-unitary quantum dynamics.  This is also consistent with recent astronomical observations of gamma-ray bursts suggesting spacetime foam does not appear to affect the propagation of free photons as certain other models of gravitational decoherence would imply \cite{Vasileiou2015}.

Nonetheless, it does not necessarily mean that realistic long-time propagations of bosonic fields suffer no quantum gravitational decoherence and dissipation either, since additional interactions will modify the argument through e.g. non-gravitational interactions and internal structures of particles and the curved nature of physical background spacetime on the cosmological scale due to cosmic expansion and on the astronomical scale due to gravity from celestial objects and dark matter.

Indeed, we have reported elsewhere how the ``no-go theorem'' for long-time decoherence represented by Eq. \eqref{tkkvsh} can be circumvented by the nontrivial Markovian quantum behaviour of bound systems, having well-defined quadrupole moments, subject to both spacetime fluctuations and other external interactions \cite{Oniga2016a}. Based on that, the aforementioned further physical considerations can also be investigated.

Finally, we remark that the present work may have far-reaching consequences for much broader real-world quantum systems consisting of a large number of particles that inevitably interact with environmental quantum fluctuations.
Especially, but perhaps not exclusively, in light of the role played by the gravitational gauge invariance \cite{Oniga2016}, it would be interesting to explore novel collective quantum effects analogous to those studied here using general transverse fields undergoing gauge invariant interactions with identical particles such as QED and beyond.

\section*{Acknowledgments}
%This research is supported by the Carnegie Trust for the Universities of Scotland (T.O.) and EPSRC/GG-Top and Cruickshank Trust (C.W.).

This research is supported by the Carnegie Trust for the Universities of Scotland (T.O.) and by the EPSRC GG-Top Project and the Cruickshank Trust (C.W.).

\end{document}